\documentclass[review,number,sort&compress]{elsarticle}

\pdfoutput=1

\usepackage{graphicx}
\usepackage{epstopdf}
\usepackage{subfigure}
\usepackage{amsmath}
\usepackage{amssymb}
\usepackage{afterpage}
\usepackage{graphics}
\usepackage{float}
\usepackage{rotating}
\usepackage{xspace}
\usepackage{dcolumn}
\usepackage{bm} 
\usepackage{lineno}

\journal{Nucl.Inst.Meth.A}

\begin{document}


\begin{frontmatter}



\title{Liquid scintillator production for the NOvA experiment}
\tnotetext[t1]{FERMILAB-PUB-15-048-ND-PPD, arXiv:1504.04035 [hep-ex]}

\author[label1]{S.~Mufson}
\address[label1]{Indiana University, Bloomington, Indiana 47405, USA}

\author[label1]{B.~Baugh}

\author[label1]{C.~Bower}

\author[label2]{T.E.~Coan}
\address[label2]{Southern Methodist University, Dallas, Texas 75275, USA}

\author[label3]{J.~Cooper }
\address[label3]{Fermi National Accelerator Laboratory, Batavia, Illinois 60510, USA}

\author[label4]{L.~Corwin}
\address[label4]{South Dakota School of Mines and Technology, Rapid City, South Dakota 57701, USA}


\author[label1]{J.A.~Karty}

\author[label5]{P.~Mason}
\address[label5]{University of Tennessee, Knoxville, Tennessee  37916, USA}



\author[label1]{M.D.~Messier}

\author[label3]{A.~Pla-Dalmau }

\author[label6]{M.~Proudfoot }
\address[label6]{Renkert Oil, Morgantown, Pennsylvania 19543, USA}


\date{\today}          

\begin{abstract}

The NOvA collaboration blended and delivered 8.8 kt (2.72M gal) of liquid scintillator as the active detector medium to its near and far detectors.  The composition of this scintillator was specifically developed to satisfy NOvA's performance requirements.  A rigorous set of quality control procedures was put in place to verify that the incoming components and the blended scintillator met these requirements.  The scintillator was blended commercially in Hammond, IN.  The scintillator was shipped to the NOvA detectors using dedicated stainless steel tanker trailers cleaned to food grade.

\end{abstract}

\end{frontmatter}



\section{Introduction}
\label{sect:intro}

NOvA\cite{MN1,nova07,nova13} is currently making precision measurements of electron-neutrino ($\nu_e$) appearance and muon-neutrino ($\nu_\mu$) disappearance. These data will help unravel unknowns in our understanding of neutrino masses and mixing.  In the standard picture of neutrinos, the three electro-weak flavor states ($\nu_e$, $\nu_\mu$, $\nu_\tau$) are mixtures of the mass eigenstates ($\nu_1$, $\nu_2$, $\nu_3$).  The flavor and mass eigenstates are related by a unitary matrix that is parameterised by three mixing angles and a charge-parity (CP) violating phase. Neutrinos are produced and detected in flavor eigenstates, but propagate as mass eigenstates.  Interference among the mass states means that a neutrino created in a definite flavor state can later be detected in a different flavor state. The oscillation probability for this process is determined by the distance the neutrino has traveled, the neutrino's energy, the mixing angles and neutrino mass splittings, and the magnitude of the CP violating phase. 

NOvA is using its near and far detectors to measure oscillation probabilities in Fermilab's NuMI (Neutrinos at the Main Injector) muon neutrino beam in order to determine the parameters of the mixing matrix.  As neutrinos travel the 810 km from the near detector at Fermilab to the far detector at Ash River, MN, through the crust of the Earth, the $\nu_e$ scatter off atomic electrons.  These interactions can either enhance or suppress the oscillation probability, depending on the parameters of the mixing matrix.  Since the effect is opposite in neutrinos compared with antineutrinos, NOvA will better our understanding of the mixing matrix parameters by comparing the oscillation probabilities of neutrinos with antineutrinos.

NOvA has been optimized to search for the rare $\nu_e$'s at Ash River that have oscillated from the $\nu_\mu$'s in the NuMI beam.  The primary design requirement for the NOvA far detector was the efficient detection of $\nu_e$ interactions at 2 GeV.  Furthermore, to minimize infrastructure costs, the plan was to operate the massive 14.4~kt far detector on the surface.  After considering several technologies, the NOvA collaboration chose to build a segmented scintillator detector.  In addition, liquid scintillator was chosen over plastic scintillator because of its significant cost advantage for massive detectors.  

This paper describes how the NOvA liquid scintillator was manufactured and delivered to the NOvA near and far detectors.  


\section{Alternative Technologies}
\label{sect:altTech}

The NOvA experiment design requires a detector technology that can
operate on the surface without appreciable dead time due to the high
cosmic-ray muon flux, that can provide good neutrino energy and
particle identification at energies of 2~GeV, and that can be built at
a cost/kiloton that enables construction of a detector with a fiducial
mass of many kilotons. These requirements favor a segmented
detector technology using liquid scintillator and disfavor
non-segmented technolgoies based on other detector mediums.

Large scale water Cherenkov detectors, like the
Super--Kamiokande experiment~\cite{bib:SuperK}, are non-segmented because the
Cherenkov effect does not yield enough light per unit path length to
enable fine segmentation while operating on the surface. 
Futher, Cherenkov detectors perform best on the relatively
simple event topologies resulting from $E<1$~GeV neutrino
interactions. Above this energy, however, the Cherenkov threshold limits the
ability to provide good hadronic energy reconstruction and particle
identification becomes complicated by multiple, often overlaping Chrenkov
rings. Super--Kamiokande also illustrates a second advantage to the
segmented design. To provide a sufficient veto against entering
particles and to provide enough distance for ring formation on the
nearest wall, Super--Kamiokande only analyzes events recorded in the 
innermost 22.5~kt of its total 50~kt detector mass. A segmented detector,
on the other hand, is sensitive over a significant fraction of its active
detector volume.  This advantage enables an overall smaller detector to achieve
the same physics. 
With the exception of the effects of the Cherenkov
threshold, these same arguments also disfavor non-segmented liquid
scintillator designs.

Other detector technologies were also considered and rejected. 
Resistive Plate Chamber (RPC) sampling calorimeters have
inferior particle identificaion efficiency for detecting $\nu_e$ events, they are more expensive to
construct, and the uninstrumented absorber regions provide paths for
comic-rays to penerate into the fiducial detector volume. They also have higher
risk of degraded performance as the gas-handling system ages. Low-Z
sampling calorimeters that use particle board as an absorber were
studied carefully because they would use building materials with
sufficient structural strength to support a massive detector. However,
they were found to have only half the electron-neutrino detection
efficiency of a segmented liquid scintillator detector and would
require a significant increase in detector mass for the same physics
reach. Liquid argon TPCs like the ICARUS detector~\cite{ICARUS} have fine resolution for charged tracks in three dimensions with an effective pixel size of $\sim 5 \times 5 \times 5$ mm$^3$.  This resolution promises enormous potential for use in neutrino physics and liquid argon TPCs appear to have the greatest efficiency for identifying $\nu_e$ interactions.  However, the largest detector operated at the time of the NOvA technology decision in 2007 (ICARUS) had about 500 tons of imaging mass and would have needed to be scaled up by about a factor of thirty to be useful in the NuMI beam intensities that were projected for NOvA.  The liquid argon calorimeter option was deemed insufficiently
mature at the time the NOvA technology choice was made in 2005.

\section{The NOvA Detector Cell}
\label{sect:detCell}

\begin{figure}[h]
\centerline{\includegraphics[width=4.5in]{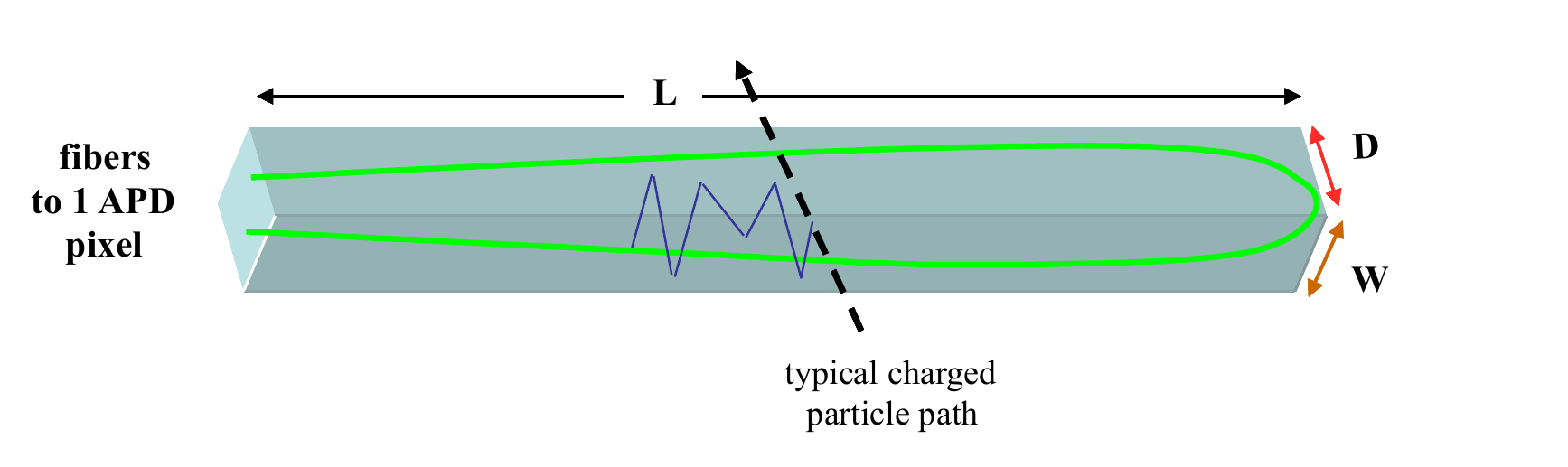}}
\caption{Schematic illustration~\cite{nova07} of a PVC cell of dimensions W$\times$D$\times$L containing liquid scintillator and a wavelength shifting fiber (green).  A charged particle incident on the cell produces scintillation photons (blue line) that bounce off the cell walls until absorbed by the fiber or lost.  The fiber routes the WLS photons to an APD pixel.}
\label{fig:NOvAcell} 
\end{figure}

The basic unit of the NOvA detector is a rigid, rectangular PVC plastic cell containing liquid scintillator and a wavelength shifting fiber.  A NOvA detector cell is schematically illustrated in Fig.~\ref{fig:NOvAcell}~\cite{nova07}.  Charged particles traverse the cell primarily along its depth (D) and scintillator light is produced along the track.  The scintillator photons bounce around in the rectangular cell of width W, depth D, and length L until they are captured by a 0.7~mm Kuraray wavelength shifting (WLS) fiber, or they get absorbed by the PVC, or they are reabsorbed by the scintillator.  The fiber is twice the length L of the cell and is looped at the bottom so the captured light is routed in two directions to the top of the cell.  Effectively there are two fibers in the cell, each with a nearly perfect mirror at the bottom so that nearly four times the light of a single non-reflecting fiber is captured.  At the top of the cell both ends of the looped fiber are directed to one pixel on a Hamamatsu Avalanche Photodiode (APD) photodetector array and the light is converted to an electronic signal.

A NOvA detector cell is made of rigid, highly reflective, titanium dioxide-loaded PVC.  There are 344,064 cells in the far detector and 20,192 cells in the near detector.  These detector  cells were commercially extruded in 16-wide modules whose interior walls are 3.3~mm thick and exterior walls are 4.9~mm thick.
They have an interior width of $W$ = 3.8~cm transverse to the beam direction, an interior depth $D$ = 5.9 cm along the beam direction, and a length of $L$ = 15.5~m.  The cell width sufficiently segments the detector so that there are many hit cells along the several charged tracks typical in 2~GeV neutrino interactions.  
The cell depth was chosen to collect enough light from the far end of the 15.5~m long cells down the wavelength-shifting fiber to trigger the APDs.  The WLS fiber captures the blue 400 -- 450~nm photons from the scintillator and wavelength shifts them to green photons in the range 490 -- 550~nm.  As the internally reflected light travels down the 15.5~m long fiber, it is attenuated by about a factor of ten, with green light preferentially surviving.  This property puts a premium on light generated in the liquid scintillator and on the use of photodetectors with good quantum efficiency in the green.  
The cell length was sized to fit on a standard U.S. domestic 53-foot semi-trailer truck.

\section{NOvA Liquid Scintillator}
\label{sect:liqScint}

NOvA liquid scintillator was formulated specifically to meet the requirements of the NOvA experiment.  Its composition intially mimicked commercially available pseudocumene-based liquid scintillators and the scintillator used in  the MACRO experiment~\cite{MACROscint}.  Modifications to the initial formulation were then developed to maximize the scintillation light.  By mass NOvA scintillator is mostly ($\sim$95\%) mineral oil solvent.  Blended into the mineral oil are a primary scintillant that generates UV light and two wavelength shifters  that convert the UV light to the wavelength range appropriate for capture by WLS fiber, the detector element that routes the light to the photodetectors.  An anti-static agent for fire safety and an antioxidant to minimize yellowing were additional components of the scintillator.

When excited by an ionizing particle, the primary scintillant pseudocumene (1,2,4-trimethylbenzene) decays by emitting photons in the range 270 -- 320~nm.  These UV photons excite the wavelength shifter PPO (2,5-diphenyloxazole) which in turn decays and emit photons mostly in the range 340 -- 380~nm, with a tail that extends to 460~nm.  In the third step in of this process, the down-converted scintillation photons excite the second wavelength shifter bis-MSB (1,4-bis-(o-methyl-styryl)-benzene) which subsequently decays to photons in the range 390 -- 440~nm, with a tail that extends to 480~nm.   Photons in the range 390-460~nm excite the wavelength shifter in the WLS fiber.

\subsection{Composition, Properties, and Mass of NOvA Liquid Scintillator}
\label{sect:composition}

Liquid scintillators with organic fluorescent compounds have a long history in particle physics~\cite{kallman,reynolds,birks}.
A mineral oil based liquid scintillator with pseudocumene as the primary scintillant and the wavelength shifters PPO and bis-MSB was used by the MACRO experiment~\cite{MACROscint}.  The composition of the NOvA liquid sctintillator is given in Table~\ref{table:composition}.  
\begin{table}[h]
\caption{The composition of NOvA liquid scintillator.}
\label{table:composition} 
\begin{center}
\begin{tabular}{|c|c|c|r|c|r|}
\hline\hline
component &purpose&mass  &  \multicolumn{1}{c|}{mass}  &mass  & \multicolumn{1}{c|}{mass} \\ 
 &&fraction &  \multicolumn{1}{c|}{(kg)}  & fraction & \multicolumn{1}{c|}{(kg)}\\ 
\hline \hline
  \multicolumn{1}{|c|}{} &\multicolumn{1}{c}{}&\multicolumn{2}{|c|}{blends: \#1, \#2 }  & \multicolumn{2}{c|}{blends: \#3 -- \#25} \\ 
\hline \hline
mineral oil &  solvent & 94.91\% & 691,179 & 94.63\% & 7,658,656\\ 
pseudocumene & scintillant & 4.98\% & 36,2677 & 5.23\% & 423,278 \\
PPO & waveshifter  & 0.11\% & 801& 0.14\% & 11,331 \\
bis-MSB & waveshifter & 0.0016\% & 11.7& 0.0016\% & 129 \\
Stadis-425 & antistatic  & 0.001\% & 7.3& 0.001\% & 81\\
Vitamin E & antioxidant & 0.001\% & 7.1 & 0.001\% & 78 \\
\hline
\multicolumn{1}{|c}{Total} & \multicolumn{1}{c}{}& \multicolumn{1}{c}{}&\multicolumn{1}{|c|}{728,247}& \multicolumn{1}{c}{}& \multicolumn{1}{|c|} {8,093,264} \\
\hline 
\hline\end{tabular}
\end{center}
\end{table}
The total volume of scintillator was manufactured in 25 separate blends of approximately 110,000 gallons each.  

As seen in Table~\ref{table:composition}, the composition of the first two blends differed from that of blends \#3 -- \#25.
The scintillator composition changed because additional wavelength shifters became available.  The NOvA far detector was originally designed to have a mass of up to 18~kt and the wavelength shifters PPO and bis-MSB for all 18~kt were purchased before construction began as the most cost-effective use of available funds.  But the detector design mass was later reduced due to financial constraints, which led to surplus wavelength shifters.  After blend \#2 a program was initiated at Indiana University to make a quantitative study of how the light yield of the scintillator was affected with additional wavelength shifters.  Increasing the pseudocumene content of the scintillator, both to generate more primary UV photons and to dissolve more solid wavelength shifter, was also part of this program.

Additional pseudocumene, however, not only increases the primary light output of the scintillator, it also lowers its flash point.  The scintillator in blends \#1 and \#2 had a flash point\footnote[1]{https://www.chilworth.com/laboratory-testing/} of 100~$^\circ$C, which meant that it was a 
Class IIIB combustible liquid, defined as a liquid with a flash point $>$~93.3~$^\circ$C.
The NOvA far detector building design (separations, exiting, fire suppression, etc.) was based on scintillator being a Class IIIB combustible liquid.  If the flash point of the scintillator were to drop below 93~$^\circ$C, the scintillator would become a Class IIIA combustible liquid with a new level of code restrictions.  Retrofitting the building for a lower flash point would have been  costly.  Further, tanker trailers transporting a Class IIIA combustible liquid would have to carry DOT Combustible Material placards with considerable increases in transportation costs.
 
As shown in Table~\ref{table:properties}, when the pseudocumene content was increased to 5.23\%, the flash point of the scintillator was lowered to 96~$^\circ$C.  Since it was time-consuming and expensive to test for even higher pseudocumene fractions, the 5.23\% pseudocumene concentration given in Table~\ref{table:composition} was adopted.  At this concentration, the WLS fiber was unaffected by the pseudocumene over the expected lieftime of the experiment \cite{MN2}.  

Once the mass fraction of the pseudocumene was established, the mass fractions of the wavelength shifters were increased independently and measurements of the light yields of the mixtures were compared with the light yield from blends \#1 and \#2. 
There was enough PPO available to increase its mass fraction in the scintillator from 0.11\% to 0.14\% and tests showed that the light yield increased proportionally up to this mass fraction.  Since PPO is easily dissolved in pseudocumene, which was the first step in the scintillator blending process, all the available PPO was included in blends \#3 -- \#25.  Additional bis-MSB, on the other hand, did not add significantly to the light yield.  
Since bis-MSB is quite difficult to dissolve in pseudocumene, the bis-MSB content was held fixed at the levels in blends \#1 and \#2.

The physical properties of the NOvA scintillator as reported on the Safety Data Sheet (SDS) are given in Table~\ref{table:properties}.  
\bgroup
\def\arraystretch{1.1}
\begin{table}[h]
\caption{Physical Properties of NOvA liquid scintillator.}
\label{table:properties} 
\begin{center}
\begin{tabular}{|l | r|}
\hline \hline
\multicolumn{1}{|c|}{property} & \multicolumn{1}{c|}{value}  \\

\hline \hline

flash point:  & \\

~~blends: \#1, \#2 & 100 $^\circ$C\\

~~blends: \#3 -- \#25 & 96 $^\circ$C\\

density (15.6 $^\circ$C) & ~~$0.862 $ g/cm$^3$  \\

water content & $\leq~50$ ppm \\

kinematic viscosity (40 $^\circ$C) & 11 cSt  \\

boiling point &$>$165 $^\circ$C \\

vapor pressure (37.8 $^\circ$C)~ & 5 mm Hg  \\

\hline \hline
\end{tabular}
\end{center}
\end{table}
The flash points for the two blends were determined by Chilworth Technology, Inc.\footnotemark[1], using the  Pensky Martens Closed Cup apparatus.  Neither blend was a Class IIIA combustible liquid.  The density given in Table~\ref{table:properties} is the mean for the 25 blend batches.  The density for each blend comes from its API gravity, the standard measure of the (inverse) specific gravity of a petroleum fluid used by the American Petroleum Institute (API).  The API gravity, which was reported on the Certificate of Analysis (COFA) by the blender for each blend batch, is determined by drawing a test sample and applying the procedures in ASTM\footnote[2]{American Society for Testing and Materials} D1298.  The mean density of the NOvA scintillator is $\rho = 0.862 \pm 0.002$~g/cm$^3$, where the error is the standard deviation of the API gravities for the 25 blends.

At a concentration of $\leq~50$ ppm, water in liquid scintillator does not affect light yield.  As reported on the COFAs, all 25 blend batches met this requirement.  This requirement on water content was also applied to the incoming mineral oil; mineral oil was rejected if it failed to meet it.  
The approved mineral oil grade, Renkert Oil Renoil 70-T, has a typical viscosity of 11~cSt at 40~$^\circ$C as determined by ASTM 445\footnotemark[2].
This viscosity led to scintillator that worked well with the NOvA pumping machines.  The boiling point and vapor pressure were required for the SDS and were measured by the blender according to standard ASTM procedures.    

To determine the emission profile for liquid scintillator, standard liquid fluorescence measurements were made with a Hitachi F-4500 Fluorescence Spectrophotometer.  A 1~cm$^2$ quartz cuvette of scintillator was illuminated from the front with 250~nm light to simulate the excitation of the pseudocumene by an ionizing particle and the spectra were viewed through the volume at 90$^\circ$ from the beam direction.  
\begin{figure}[h]
\centerline{\includegraphics[width=3.1in]{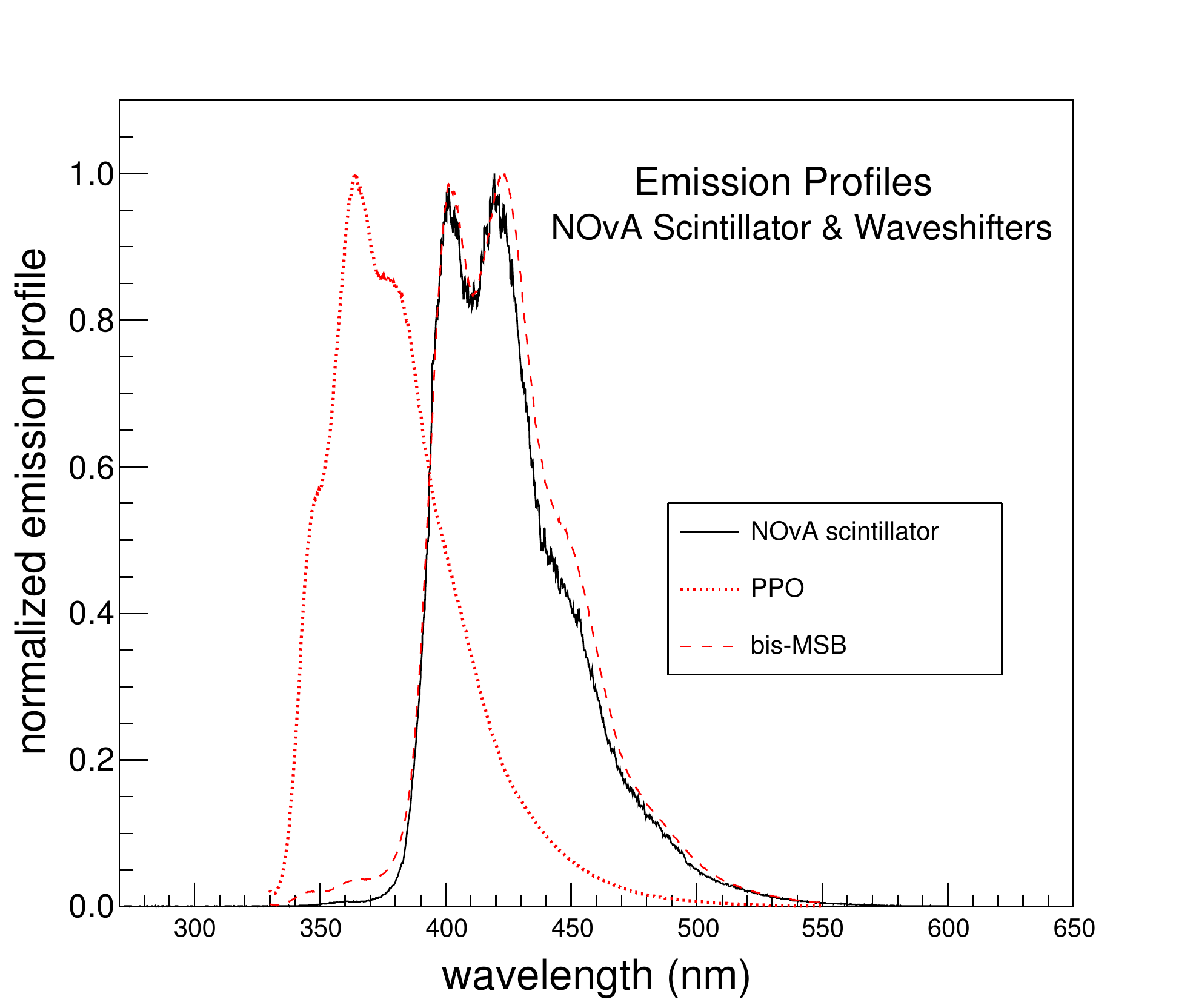}}
\caption{The normalized emission profiles for blended NOvA scintillator and the wavelength shifters PPO and bis-MSB dissolved in toluene.  The scintillator peak was normalized to 1.  To show the relative contributions of the wavelength shifters to the scintillation light, the emission profiles for the wavelength shifters were also normalized to 1.  The scintillation light is dominated by the emission from bis-MSB.}
\label{fig:emissionProfiles} 
\end{figure}
The resulting normalized scintillator emission profile is shown as the ``NOvA scintillator trace'' in Fig.~\ref{fig:emissionProfiles}.  As a check, excitation of the NOvA scintillator sample with light in the range 270--300~nm, the output spectral range of pseudocumene scintillation, resulted in the same emission profile.  For quality control purposes, emission spectra were regularly obtained for both PPO and bis-MSB dissolved in toluene using 320~nm front-side excitation.  These spectra were also viewed from the front side.  Their normalized emission profiles are superposed in Fig.~\ref{fig:emissionProfiles}.  The distinct two peak emission profile of the scintillator sample makes it clear that the emission from NOvA scintillator is dominated by emission from bis-MSB.

The volume and mass of the scintillator in the near and far detectors are given in Table~\ref{table:mass}.  
\begin{table}[h]
\caption{The mass and volume of liquid scintillator in the NOvA near and far detectors.}
\label{table:mass} 
\begin{center}
\begin{tabular}{|c | c|c|}
\hline \hline
 \multicolumn{1}{|c|}{detector}& volume & mass \\
\multicolumn{1}{|c|}{}&  (gal) & (kg) \\
\hline \hline
near detector & 40,141 & 130,672\\
far detector & 2,674,041&  8,690,929 \\
\hline
Total &2,714,182 & 8,821,511\\
\hline \hline
\end{tabular}
\end{center}
\end{table}
The volume of scintillator comes from the records of the volume pumped into the near and far detectors.  Since the density given in Table~\ref{table:properties} is measured at 15.6~$^\circ$C (60~$^\circ$F), which is different from the actual scintillator temperature in the detectors, the density requires a temperature correction factor for the conversion from volume to mass.  These are standard corrections and they are found in ASTM D1250-08\footnotemark[2].  The temperature of the liquid scintillator at Ash River is 20.6~$^\circ$C; 
the density of the far detector scintillator is $\rho = 0.859 \pm 0.002$ g/cm$^3$.  The temperature of the liquid scintillator at near detector at Fermilab is 18.3~$^\circ$C; 
the density of the near detector scintillator is $\rho = 0.860 \pm 0.002$ g/cm$^3$.


\subsection{Scintillator Blending Operations}
\label{sec:blendingOperations}

The NOvA liquid scintillator was blended at Wolf Lake Terminals\footnote[3]{http://www.wolflakeinc.com} in Hammond, IN under the supervision of Renkert Oil, LLC\footnote[4]{http://www.renkertoil.com/default.html}.  The scintillator production operation at Wolf Lake is shown schematically in Fig.~\ref{fig:blendingSchematic}. 
\begin{figure}[h]
\centerline{\includegraphics[width=5.4in]{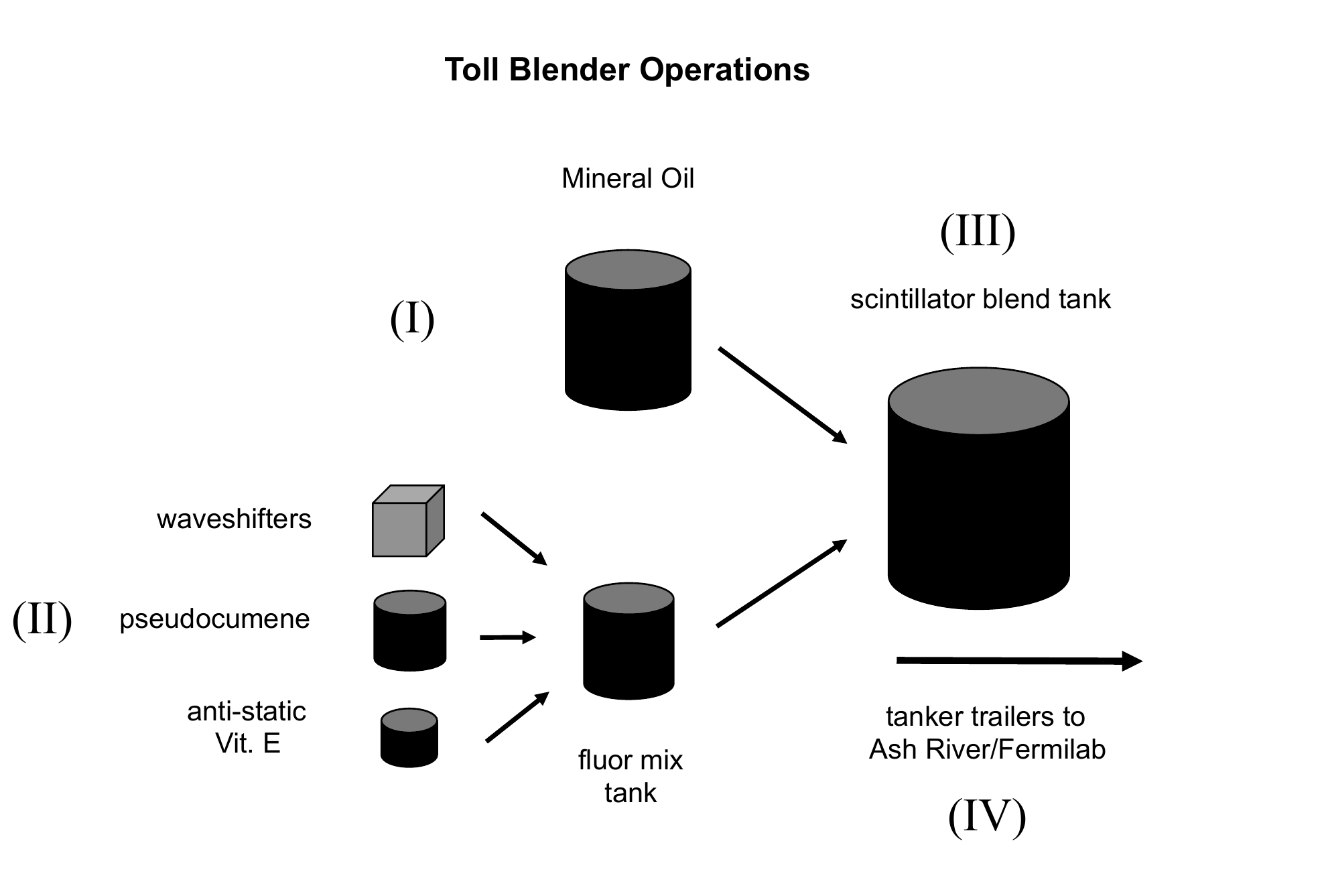}}
\caption{Schematic of blending operations for NOvA liquid scintillator at Wolf Lake Terminals.   (I)~Mineral oil was brought to Wolf Lake Terminals from storage/rail car and pumped into the scintillator blend tank.  (II)~Fluors and additives were mixed at the NOvA mixing station and pumped into the blend tank.  (III)~The scintillator was blended with a Pulsair$^{TM}$ mixing system that uses bubbles of N$_2$ gas and then (IV) shipped to the NOvA detectors by tanker trailer.}
\label{fig:blendingSchematic} 
\end{figure}

(I)~Mineral oil was typically brought to Wolf Lake from a dedicated 600,000 gallon storage tank at Westway Terminals\footnote[5]{http://www.westway.com}, Hammond, IN.  This tank was a buffer against the risk of delaying the production schedule because shortages made mineral oil unavailable or because mineral oil experienced a dramatic, unanticipated price increase that the Project would choose to wait out.  The tank was first carefully cleaned and then coated with Carboline Plastite 3070\footnote[6]{http://www.carboline.com}.  This coating was subjected to accelerated aging tests with NOvA scintillator (heating Al-coated samples in scintillator to 32~$^\circ$C for months) and it was found to have minimal impact on scintillator transparency, light yield, and chemical composition.  When feasible,  rail cars of mineral oil were brought directly to Wolf Lake from the mineral oil production facility in Lousiana to lower costs.  

(II)~The fluors (pseudocumene, PPO, \& bis-MSB), the anti-static additive, and the Vitamin E were mixed into a ``fluor blend'' as a first step in the blending process at a specially designed, dedicated NOvA mixing station at Wolf Lake.  The pseudocumene was first loaded into a 6,500 gallon stainless steel fluor mix tank.  About 300 gallons of pseudocumene were pumped from this mixing tank into a stainless steel vessel where a 0.5~kg container of bis-MSB was added and mechanically stirred for 30 minutes.  When the bis-MSB was fully dissolved, the mixture was pumped back into the mixing tank.  The process was repeated until all the containers of bis-MSB had been dissolved.  Bis-MSB is not easily dissolved in pseudocumene and the whole process took several hours.  Once the bis-MSB had been dissolved,
the PPO was added to the blend.  Since the solubility of PPO in pseudocumene is much higher than that of bis-MSB, the 30~kg containers of PPO were successively added to the mixing vessel at once and stirred for 20 minutes.  Finally the anti-static agent and the Vitamin E were added.  The preparation of the fluor mix took approximately 12 hours.  When all components of the fluor blend had been combined, the mixture was pumped into the fluor mix tank and circulated in the tank for several hours to achieve a homogeneous solution.  

(III)~The fluors and mineral oil for each blend were pumped into a 120,000 gallon scintillator blend tank that had also been thoroughly cleaned and lined with Carboline Plastite 3070.  The scintillator was then mixed with large bubbles of dry N$_2$ gas from a Pulsair industrial tank mixing system\footnote[7]{http://www.pulsair.com}.  Since O$_2$ has long been known to quench scintillation light\cite{o2quenching}, sparging with N$_2$ gas has the advantage of purging the O$_2$ from the scintillator, thereby yielding scintillator brightened to its maximum.  
There were two blend tanks used in the NOvA blending production.  While one tank was used to blend a batch of scintillator, the second was used for loading the tanker trailers for scintillator transport.  There were 25 blending cycles required for the total NOvA liquid scintillator production, each of which had a typical blend volume of 111,500 gal.  

(IV)~The blended scintillator was transported to NOvA detectors at Ash River and Fermilab by dedicated stainless steel 7,000 gallon tanker trailers.  These tanker trailers were first cleaned to food grade and then tested for cleanliness by looking for contamination with a clean mineral oil rinse.  Once in service, the tanker trailers were not cleaned again.  There were 410 tanker trailer loads required for the entire volume of liquid scintillator.

\subsubsection{Mineral Oil (I)}

The main component and the primary cost-driver of NOvA liquid scintillator was mineral oil.  The mineral oil used in blending NOvA liquid scintillator was the technical grade white mineral oil\footnote[8]{21CFR178.3620} Renoil 70-T obtained from Renkert Oil, LLC.  It was chosen from seven candidates based on a competitive bidding process.  The bid oils were judged on both cost and technical criteria.

The most important technical requirement on the mineral oil is its attenuation length.  The attenuation length requirement was established by a Monte Carlo simulation in which a NOvA detector cell as shown in Fig.~\ref{fig:NOvAcell} was modeled as a PVC extrusion containing the design fraction of anatase TiO$_2$ \cite{nova07}.  Through the extrusion ran a loop of 0.7~mm diameter WLS fiber that captured scintillation photons and transmitted them to an APD photodetector at one end.  In this simulation photons were released randomly throughout the volume and propagated as particles until they were captured by the fiber or lost to absorption in the walls or reabsorption in the scintillator.  At the walls the photons were reflected either specularly or diffusely, according to the experimental fractions for anatase TiO$_2$.  

The results of the simulation showed that the attenuation length requirement on the NOvA liquid scintillator was rather modest.  The mean pathlength traveled by a photon before absorption by the WLS fiber was only 0.4~m, although the tail extended out to much longer pathlengths.  However, virtually all of the photons were collected within 1.5~m.  The search for mineral oil appropriate for NOvA scintillator then became an experimental program to find mineral oil that resulted in scintillator with an attenuation length $>$~1.5~m when blended with baseline fluors.  To account for the many simplifications in the simulation, the requirement adopted was more conservative: scintillator was required to have an attenuation length $>~$2~m.

In this experimental program 10 mineral oils with attenuation lengths ranging from 2.5~m to 15~m were blended into scintillators.  The goal was to find which of these mineral oils resulted in scintillators that would meet the  2~m attenuation length requirement.  The mineral oils were blends made with different fractions of Penreco's Parol with an attenuation length of 2.5~m and Renkert Oil's Renoil 70-T with attenuation length 15~m.  The IU spectrophotometer apparatus used to measure the attenuation lengths of the mineral oil blends and the scintillator blends is described in~\cite{petrakis}.  Results showed that mineral oil with an attenuation length $>$3.25~m could be blended into scintillator that meets the NOvA attenuation length requirement.

NOvA needed a reliable, accurate, and rapid method of testing the attenuation length for a very large number of samples of mineral oil and scintillator.  IU spectrophotometer measurements were too time-consuming. Rail cars and barges of mineral oil had to be tested prior to acceptance; scintillator batches had to be tested after blending; tanker trailers of scintillator had to be tested prior to shipping to Ash River or Fermilab, as well as at arrival in case the scintillator had been contaminated during transport.  The device used in production for all these tests was a Lovibond PFX880 tintometer\footnote[9]{http://www.lovibondcolour.com}, which (in this application) measured the fractional transmission of a standard beam of light at 420~nm.   To guard against systematic errors affecting the tintometer results at the many stations where it was used (IU, Fermilab, Wolf Lake, Ash River), all sample measurements were made with respect to an optical glass standard purchased from Edmund Optics\footnote[10]{http://www.edmundoptics.com}. 
\begin{figure}[h]
\centerline{
\includegraphics[width=2.67in,height = 2.3in]{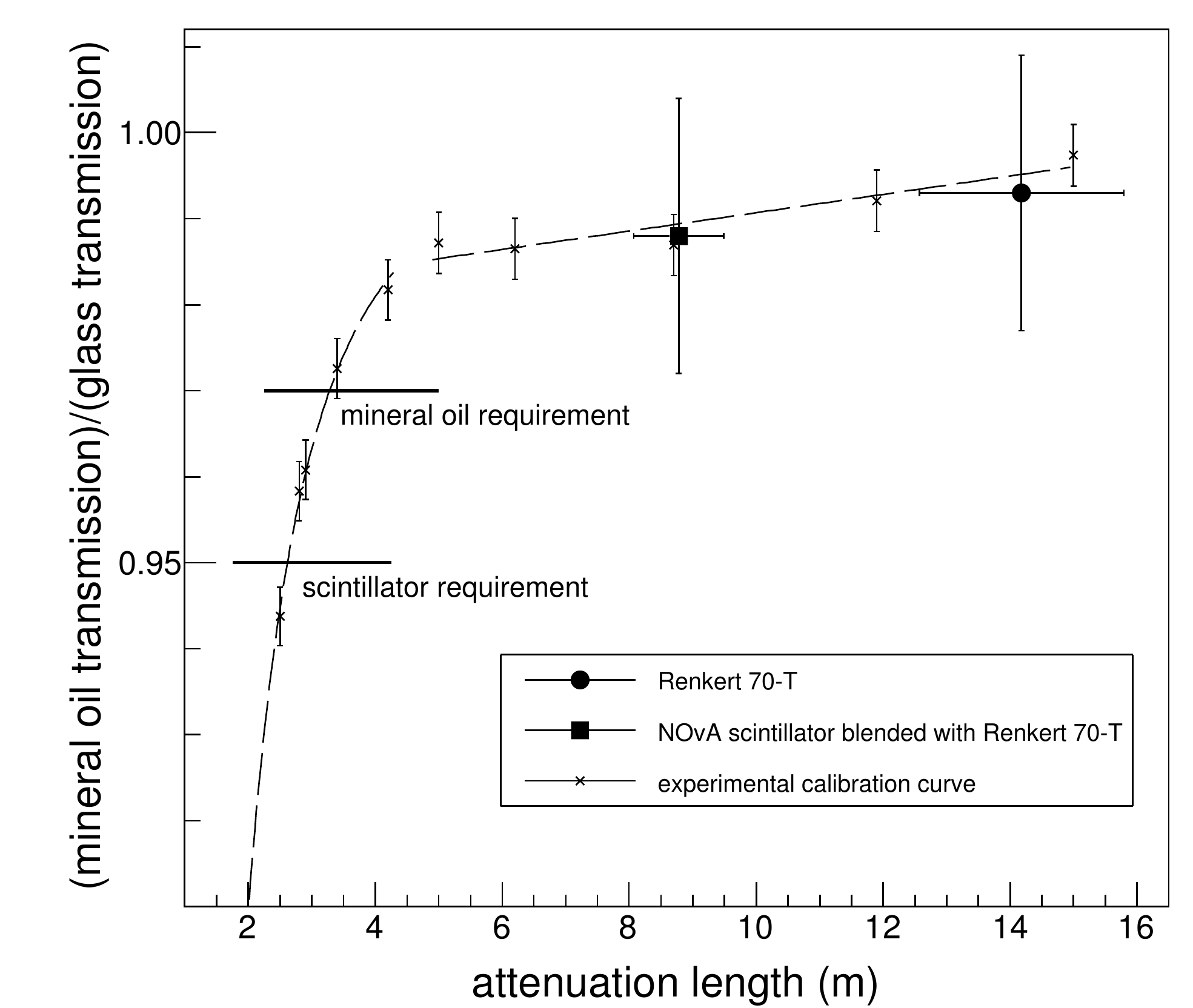}
\includegraphics[width=2.72in,height = 2.5in]{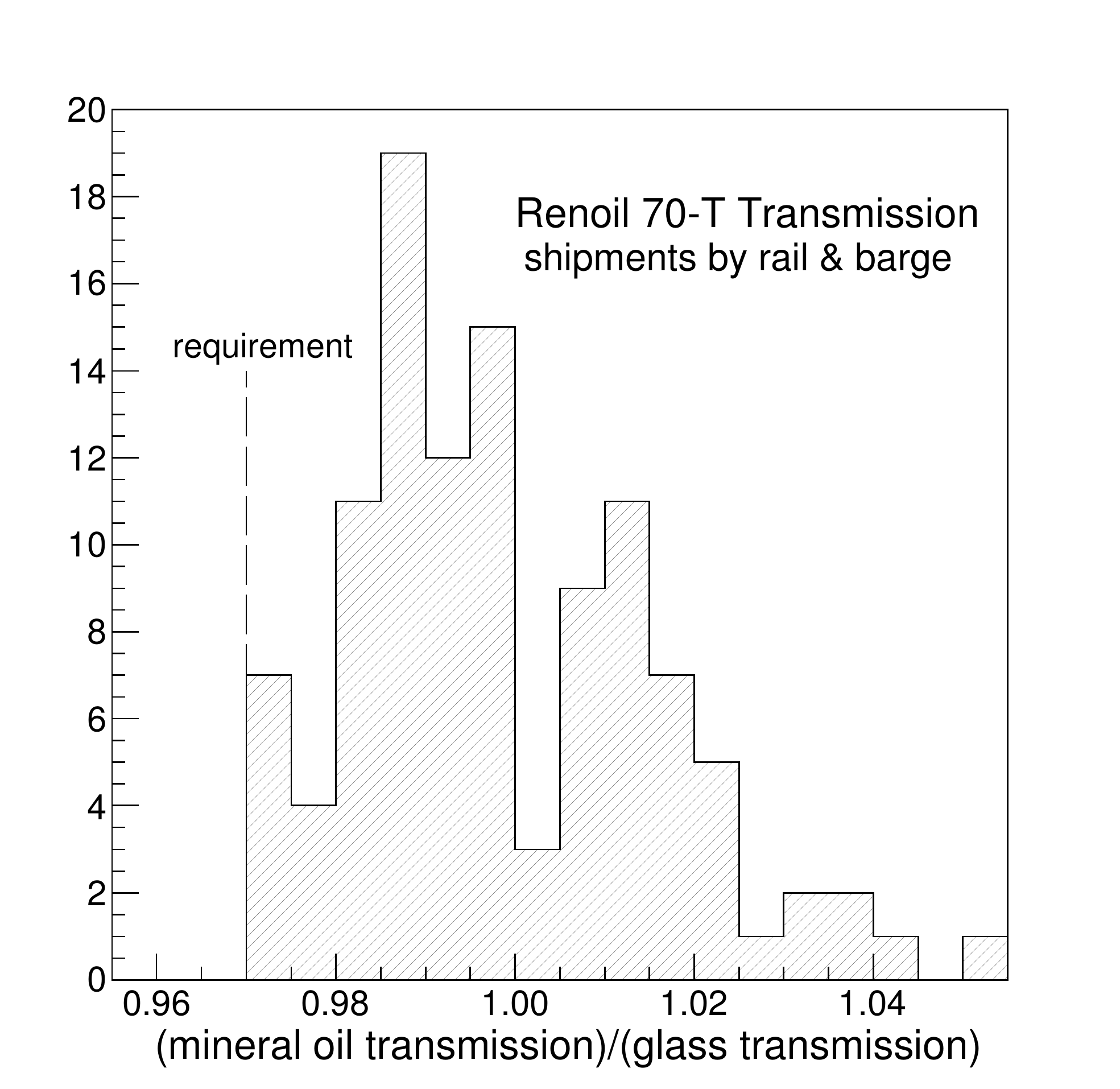}
}
\caption{{\it Left:} The calibration curve for the conversion of tintometer measurements to attenuation length, as measured by the IU spectrophotometer, based on 10 mineral oil standard blends.  The means for the mineral oil used to blend the scintillator and the blended scintillator are shown.  {\it Right:} Histogram of tintometer measurements of Renoil 70-T for all the rail cars or barge loads of mineral oil used in blending NOvA scintillator.  The tintometer acceptance requirement is marked.}
\label{fig:attnLenTrans}
\end{figure}
The calibration curve for the conversion of tintometer measurements to attenuation length, from repeated measurements of the mineral oil standards, is shown in Fig.~\ref{fig:attnLenTrans}.  
The calibration curve has two regimes which reflect the 6$''$ length of the tintometer sample cell.  Fits to these regimes have been overlayed on the calibration data.  Up to $\sim$4.5~m, the tintometer can accurately measure the attenuation length.  For attenuation lengths $>$~5~m, the tintometer has much less discrimination power.  In that regime, however, the mineral oil and scintillator meet requirements.

Using the mineral oil standards and the scintillator composition for blends \#1 and \#2 in Table~\ref{table:composition}, it was found that mineral oil with an attenuation length of $>$3.25~m results in scintillator with attenuation length $>$2~m.  These experiments established the mineral oil transparency requirement, tintometer $>$~0.97, and the scintillator transparency requirement, tintometer $>$~0.95, as shown on Fig.~\ref{fig:attnLenTrans}.  (Numbers were rounded to two decimal places.)

Also shown in Fig.~\ref{fig:attnLenTrans} are the mean attenuation length and its standard deviation for 5 different samples of Renoil 70-T obtained from Renkert Oil during the experiments.  The mean and standard deviation for the tintometer measurements were taken from the right panel of Fig.~\ref{fig:attnLenTrans}, which shows the tintometer readings for all mineral oil shipments used in blending NOvA scintillator.  For the blended scintillator, the mean and its standard deviation were measured for scintillators blended from the mineral oil samples.  The mean and standard deviation for the tintometer measurements were taken from Fig.~\ref{fig:trans_lightYield}.

\subsubsection{Fluor Blend (II)}

\medskip

1. Pseudocumene

\medskip

\noindent  The primary scintillant in NOvA liquid scintillator is pseudocumene (PS).  The distribution of its scintillation light in response to ionizing particles is strongly peaked between 285-290 nm with a FWHM of $\approx$~40~nm.  The distribution of scintillation photons rises sharply to its peak from 270~nm and then falls off more gradually to zero at $\sim$~350~nm~\cite{nova07}.  The pseudocumene for the NOvA scintillator was purchased from the Chinese chemical company Aquachem Industrial Limited\footnote[11]{http://www.aquachemi.com/Product/Petrochemicals/PCM.html}.  The primary requirements on the pseudocumene were purity ($>$ 98.1\%) and low sulfur content ($<$ 1 $\mu$g/g).  The purity measurements on the manufacturer's COFA were compared with gas  chromatography-mass spectrometry (GC-MS) measurements at the Indiana University Mass Spectrometry Facility (IUMSF)\footnote[12]{http://msf.chem.indiana.edu} before acceptance.  All 25 ISO tankers of pseudoucumene (one $\sim$5,700 gal ISO tanker per scintillator blend) had measured purities $>$ 99\%.  All deliveries of pseudocumene met the sulfur content requirement as reported in the manufacturer's COFAs.

Measurements of the pseudocumene mass fraction in fluor blend samples proved to be challenging.  Repeated measurements of the same sample showed that errors in GC-MS determinations of the PS mass fraction were typically $\sim$0.7\%.  In a separate set of tests, 10 samples of scintillator were blended with a known PS mass fraction of 4.468\%.  All GC-MS measurements were low by an average systematic offset of 1.2\%.  

\medskip

2. PPO and bis-MSB

\medskip

\noindent The wavelength shifters were purchased from Curtiss Laboratories in Bensalem, PA.  The PPO absorption cross section has a peak at 300~nm, a FWHM of 50~nm, and so absorbs all the scintillation light from PS.  The light is reemitted with a spectrum having a broad peak from $\sim$350--400~nm and a long tail extending out to $\sim$500~nm, as seen in Fig.~\ref{fig:emissionProfiles}.  The bis-MSB has an absorption peak at 345~nm and a FWHM 70~nm.  The light is reemitted with a double-peak spectrum with a tail extending to 550~nm, also seen in Fig.~\ref{fig:emissionProfiles}.  Light yield tests showed that scintillator with both PPO and bis-MSB gave a stronger scintillation signal than with PPO or bis-MSB alone.  

The requirements on the wavelength shifters for several properties, including purity ($>$99.0\%), melting point, and transmittance in toluene, were reported on the COFAs from the manufacturer.  The melting point and transmittance in toluene were verified at Fermilab.  Nuclear magnetic resonance (NMR) and infrared spectra for all lots of wavelength shifters were recorded at Northern Illinois University to check for contaminants such as solvents, moisture, unreacted materials and by-products.  No contaminants were identified in any delivery.  

Measurements of the wavelength shifters in the fluor blend samples were made by high-performance liquid chromatography (HPLC) at the IUMSF.  Repeated measurements of the same sample showed that statistical errors in the measurements were $\sim$0.5\%.  No systematic offsets were observed.

\medskip

3. Additives

\medskip

\noindent The antistatic agent Stadis-425 was obtained from Innospec Fuel Specialities\footnote[13]{http://www.innospecinc.com}.  It is a common fuel additive used to increase the conductivity of nonconducting fluids like NOvA scintillator.   Nonconducting NOvA scintillator presents a fire hazard due to the potential buildup of static charge when loaded into the 15.5~m  long NOvA extrusions.  The technical requirement for conductivity was taken from the recommendations given by the National Fire Protection Association (NFPA).  NFPA safe practices dictate that the scintillator be made ``semi-conducting'', which is defined as ``possessing a conductivity at least 100 picosiemens/meter''\footnote[14]{NFPA 77, http://catalog.nfpa.org/NFPA-70-National-Electrical-Code-C3315.aspx}.  The requirement was increased to 125 pS/m to guarantee that small errors in measuring out small quantities of Stadis-425 did not compromise the conductivity of the scintillator during its production.  This requirement could be met by adding 3 ppM of Stadis-425 to the scintillator.  This concentration of Stadis-425 did not affect the transmission or light yield of the scintillator.  

NOvA needed a reliable and efficient method of measuring the conductivity of blended scintillator to be confident that the requirement was met.  Conductivity measurements were made with an Emcee Electronics Model 1152 Digital Conductivity Meter\footnote[15]{http://www.emcee-electronics.com/product/model-1152-digital-conductivity-meter/}.

The Vitamin E to prevent yellowing was supplied by Renkert Oil.  The concentration of Vitamin E used in the NOvA blend was the same as that added by MACRO~\cite{MACROscint}.

\medskip

4.  Quality Control for the Fluor Blend

\medskip

\noindent After a fluor blend was mixed, a sample was sent to Indiana University for QC analysis.  The blend was mixed with the mass fraction of mineral oil needed to match the scintillator composition in Table~\ref{table:composition}.  In practice it proved difficult to accurately blend the small batches of scintillator needed for these tests so individual measurements of the mass fractions of the components were not reliable.  The approach taken was to measure the ratios of PPO to PS and bis-MSB to PS which canceled out the errors in the mass of mineral oil added to the samples.  So long as the ratios were correct, the addition of the proper amount of mineral oil for a 110,000 gallon blend would lead to the correct scintillator composition.  Small errors in the mass of mineral oil added would not affect the composition.  Fig.~\ref{fig:PPO_bis_PS} shows the results of measuring the ratios for all scintillator blends.  
\begin{figure}[h]
\centerline{
\includegraphics[width=2.55in]{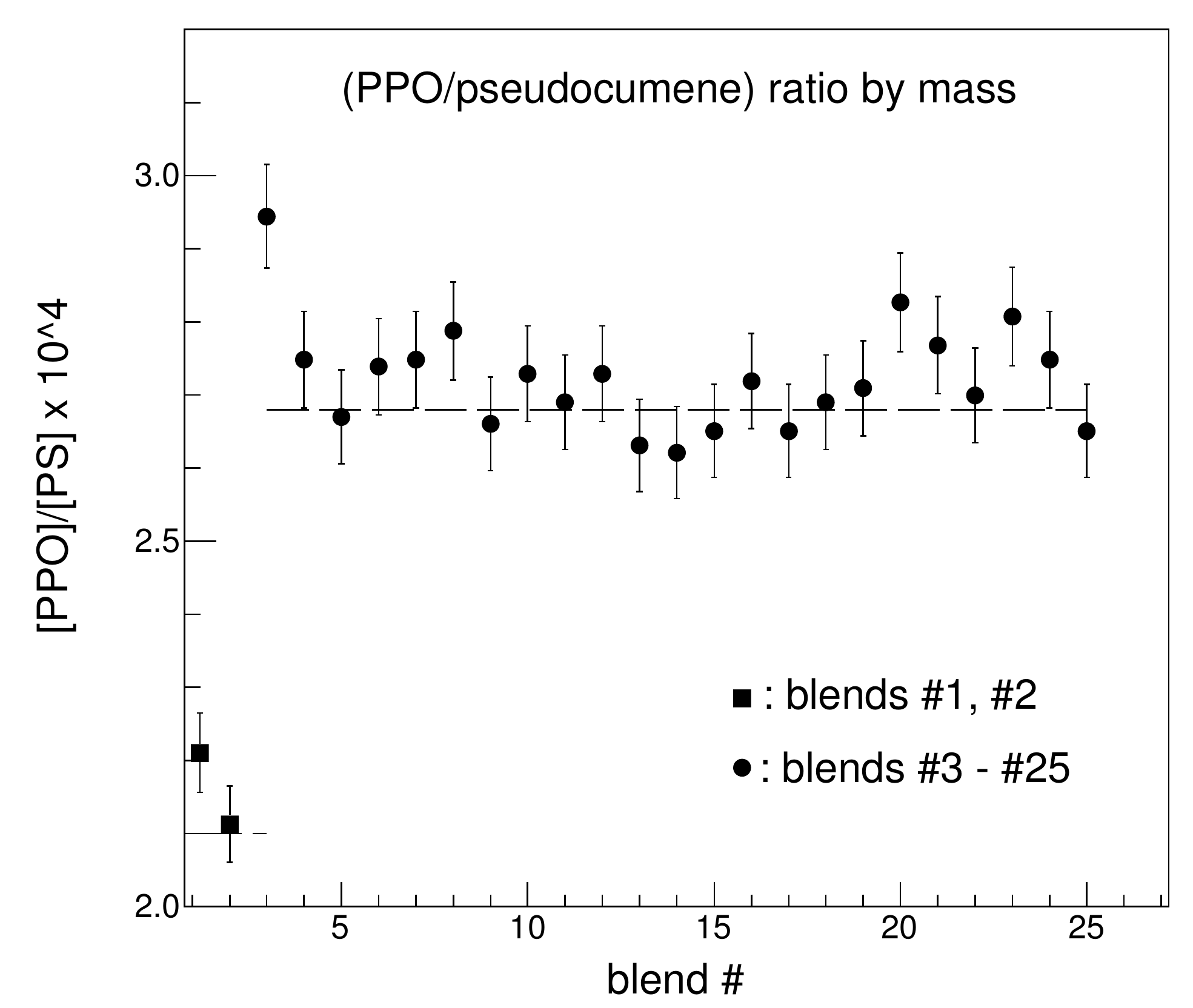}
\includegraphics[width=2.55in]{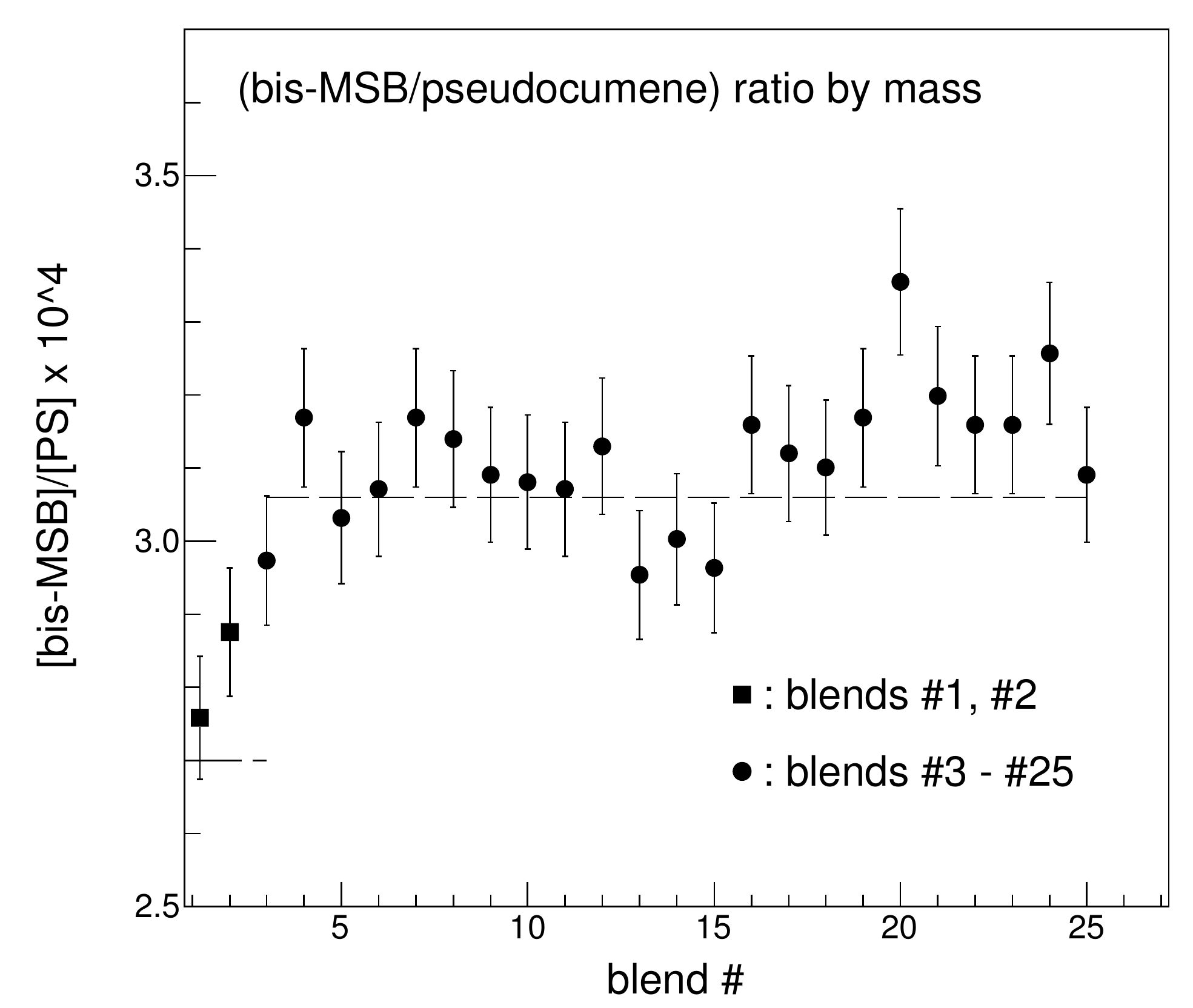}}
\caption{{\it Left:} The ratio of the wavelength shifter PPO to pseudocumene by mass for all blends of NOvA scintillator.  The dashed lines are the expected baseline ratios.  {\it Right:} The ratio of the wavelength shifter bis-MSB to pseudocumene by mass for all blends of NOvA scintillator.  The dashed lines are the expected baseline ratios. }
\label{fig:PPO_bis_PS}
\end{figure}
The mass fractions in the ratios were corrected for the purity of the PS, PPO, and bis-MSB as reported in the COFAs.  The error bars were determined from the statistical errors in the determination of the mass fractions propagated into the ratios.  The ratios are biased high.  This bias is consistent with the systematically low measurements of the PS mass fraction.  The mass fraction of PPO in blend \#3 was known to be high.  More light from the additional PPO was not considered a problem.  All blends were judged to be acceptable.  

\subsubsection{Quality Control for Blended Scintillator}

\medskip

1. Transmission

\medskip

\noindent Blend-by-blend tintometer measurements of the scintillator transmission for all 25 blend batches are shown in the left panel of Fig.~\ref{fig:trans_lightYield}.  All blends met the transmission requirement shown in Fig.~\ref{fig:attnLenTrans}.
\begin{figure}[h]
\includegraphics[width=2.35in,height=1.95in]{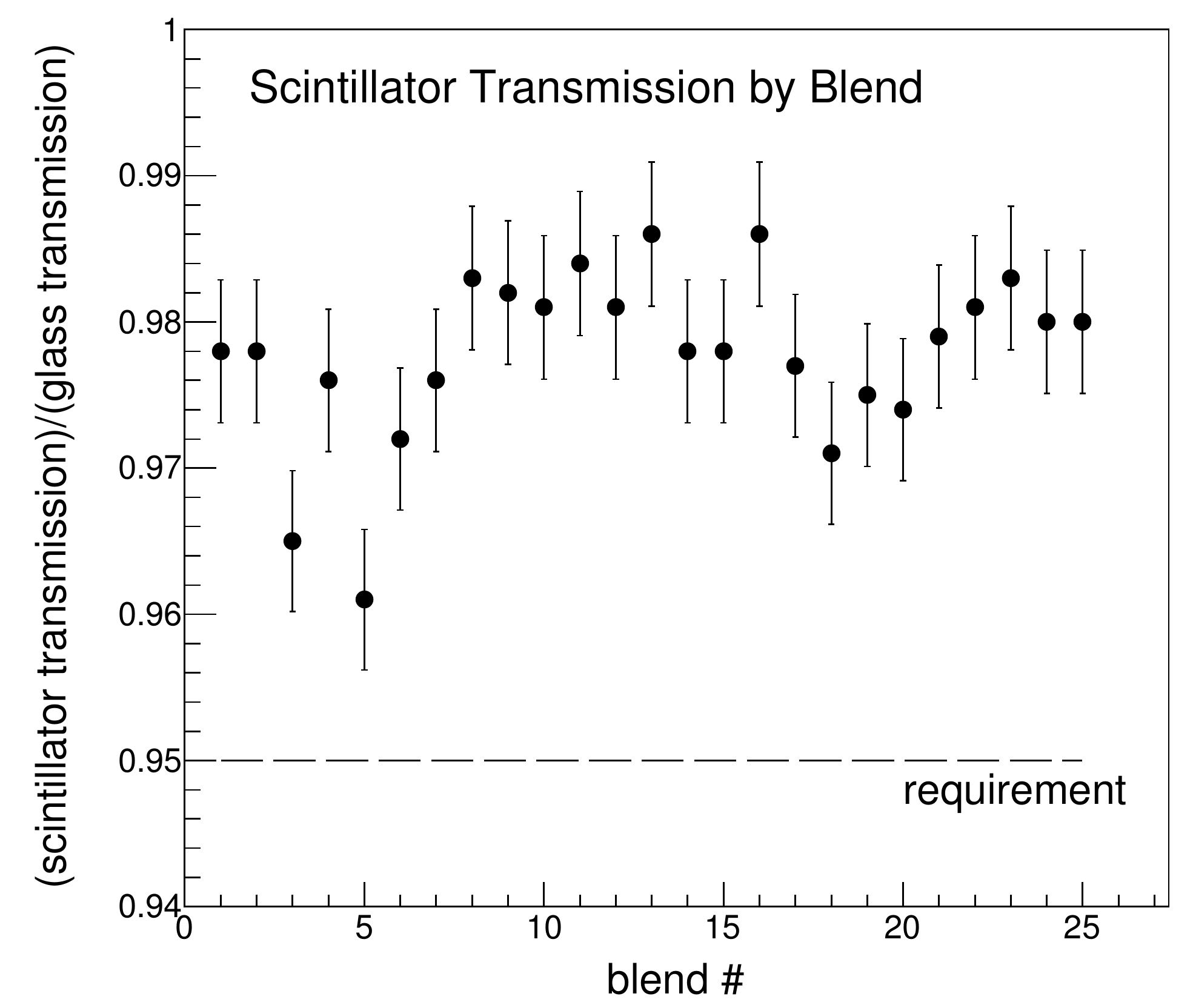}
\includegraphics[width=2.35in,height=1.95in]{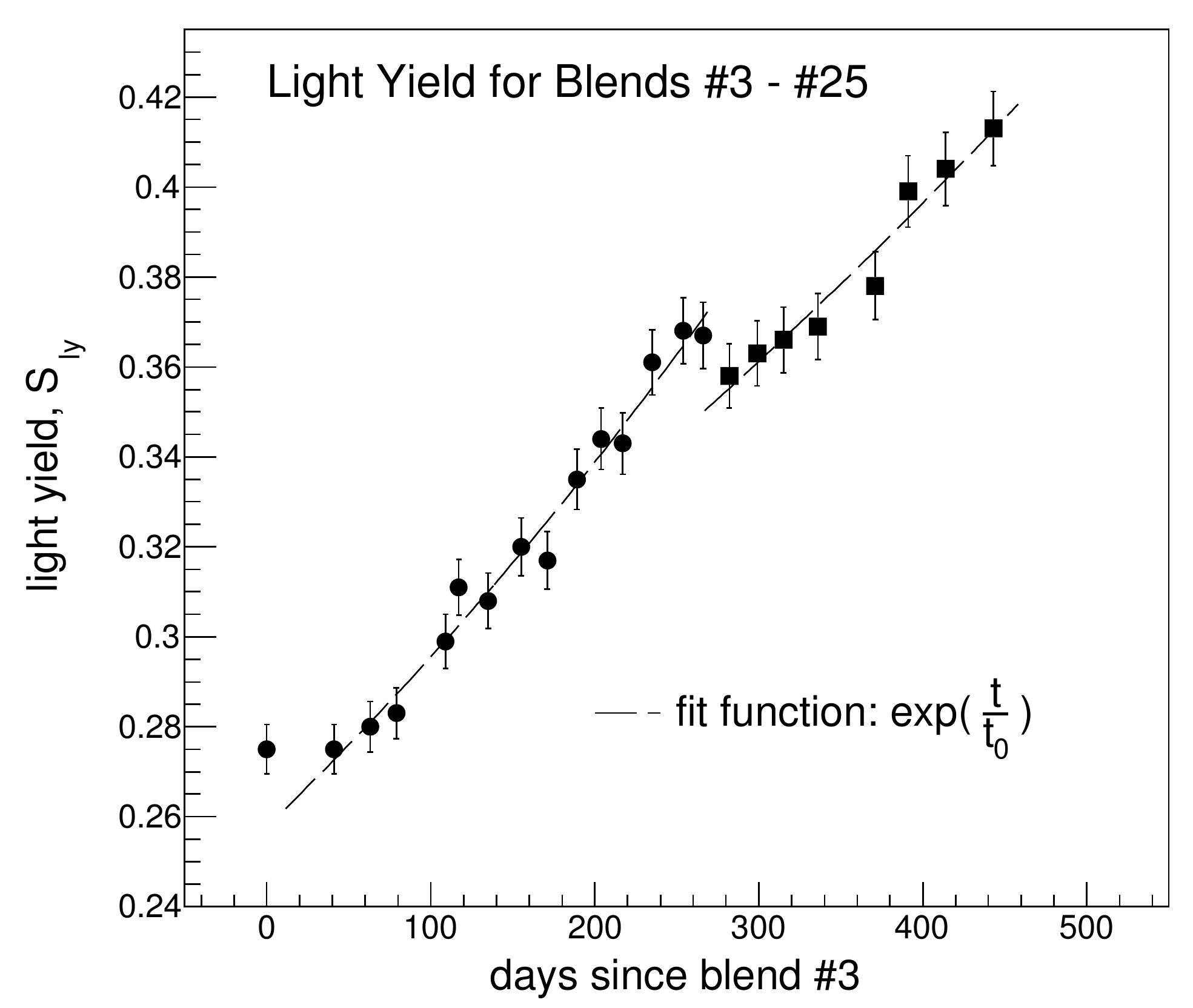}
\caption{{\it Left:}  Scintillator transmission measurements for all 25 blend batches.  All blends met the transmission requirement.  {\it Right:} Results of the light yield test for blends \#3 -- \#25.  The apparent rise in the light yield is due to radiation damage to the plastic scintillator from the $^{241}$Am $\alpha$ source used in the test.   }
\label{fig:trans_lightYield}
\end{figure}

\medskip

2. Light Yield

\medskip

\noindent The light yield test was designed to measure whether the light from a blend batch of scintillator met the standards required to reach NOvA's science goals.  Extensive testing in the lab and in prototypes showed that the scintillation light with the baseline scintillator composition in Table~\ref{table:composition} was sufficient to meet this requirement, as long as no contaminants are introduced that quench light.  For blends \#1 and \#2, the light yield was comparable to Bicron BC-517P saturated with air, which has 21\% of the light yield of anthracene.\footnote[16]{http://pe2bz.philpem.me.uk/Comm01/}  The measurements were made with respect to a single batch of BC-517P less than a few months old.  With additional pseudocumene and PPO, the light yield for blends \#3 -- \#25 increased by approximately 10\% relative to the baseline composition.  These measurements were made by comparing laboratory blends with baseline scintillator.

Usually a light yield test for production would compare the light yield of a scintillator batch with a light yield standard.  As described below, however, a suitable light yield standard was not found.  Instead, the test implemented during production relates the light yield of a scintillator blend batch directly to its composition.  It measures whether the light yield, and therefore the composition, remains consistent from batch to batch.  This consistency confirms that the scintillator was blended correctly and guards against the introduction of contaminants that quench scintillator light.  For a blend batch  to be approved, its light yield had to be consistent with previous blends and its chemical composition had to meet requirements based on independent chemical tests.                                                                                                                                                                                                                                                                                                                                                                                                                                                                                                                                                                                                                                                                                                                                                                                                                                                       

The search for a light yield standard initially was an attempt to identify a liquid scintillator standard whose light yield could be tested in the same way as the blended scintillator to minimize systematics.  This search included commercially blended scintillators and laboratory blended scintillators.  Batch to batch variations in both commercial and laboratory scintillators made their light yields too inconsistent to use as a standard.  
Since the liquid scintillator yellows and its light yield degrades when exposed to the atmosphere over many months to years, storing a single blend of scintillator to use as a standard throughout the production was believed to be too inaccurate, even after N$_2$ gas was bubbled through it as a way to restore the light yield.
When production started, a plastic scintillator standard was adopted as an alternative to liquid scintillator.  After the first few blends, however, it became clear that the plastic scintillator standard was not perfoming reliably.  
With production at typically 25,000 to 50,000 gal/wk, time was not available to make further attempts at identifying a light yield standard.  Consistency from batch to batch along with the tests of chemical composition were considered sufficient.

The light yield apparatus consisted of a 100~ml sample of scintillator in a jar keyed to a frame that housed a 3'' Burle~S83049F PMT biased at 1040~V.  The keyed scintillator sample jar always sat on the PMT in the same orientation.   The sample jar was irradiated from the side with a 10~$\mu$Ci $^{137}$Cs $\gamma$~source, which provided a signal, $S_{CE}$, from Compton scatters at the Compton edge.  A 1~$\mu$Ci $^{241}$Am $\alpha$~source sat on the PMT in an aluminum can with a plastic scintillator window.  The light signal from the irradiated plastic scintillator provided a fiducial signal, $S_\alpha$, that was used to remove systematics due to electronics drifts.  The PMT was read out by an ORTEC EASY-MCA\footnote[17]{http://www.ortec-online.com/Solutions/multichannel-analyzers.aspx} that sorts the events into a histogram as function of pulse height or energy.  

Let $S_{CE}$ = Compton edge signal from the scintillator sample and $S_\alpha$ = fiducial signal from the $\alpha$~source.  The statistic, $ S_{ly}$, used to measure the light yield is given by
\begin{equation}
S_{ly} = S_{CE}/S_\alpha,
\end{equation}
which can be rewritten 
\begin{equation}
S_{ly} = K \, N_{scint}/N_{pl},
\end{equation}
where $N_{scint}$ is the number of scintillator molecules in the 100~ml sample irradiated by the $\gamma$~source and $N_{pl}$ is the number of plastic scintillator molecules irradiated by the $\alpha$~source.  In this equation, $K$ collects the experimental constants, including the activities of the radiological sources and the cross sections for the interactions. 
The MCA response function and the sampling time of the MCA, cancel out in the ratio.  Consequently, this metric removes systematics due to the measurement apparatus and electronics drifts.  The measurements of $S_{ly}$ for blend batches \#3 -- \#25 are shown in Fig.~\ref{fig:trans_lightYield}.   Light yield measurements from blend batches \#1 and \#2 were consistent with one another but the electronics setup and PMT voltage setting were different from those used in blends \#3 -- \#25.  Measurements from these blends were not directly comparable to the later blends without an absolute standard to tie them together.

The MCA data were analyzed with standard ROOT fitting algorithms.  The Compton edge was identified as the inflection point in the abrupt fall-off of the Compton signal from the scintillator.  The fiducial $\alpha$ signal was seen as a distinct peak offset from the Compton signal and was fit with a gaussian function.  $S_{ly}$ was calculated as the ratio of channel of the inflection point to the channel of the peak of the gaussian.  Since $S_{ly} \propto N_{scint}$, the light yield test in principle determines whether the scintillator composition was correct by measuring a key performance characteristic, its light yield.  When blended correctly, $S_{ly}$ was expected to remain approximatey constant, which would make the light yield test an independent confirmation of the chemical tests of scintillator composition.  In practice, $ S_{ly}$ continuously rose during the entire period of scintillator production, as seen in Fig.~\ref{fig:trans_lightYield}.  When examined, it became clear that the scintillator plastic used as a window on the $\alpha$ source was suffering radiation damage.  At blend \#18, the plastic scintillator was replaced (and the PMT voltage adjusted) but the rise continued.  

Fig.~\ref{fig:trans_lightYield} suggests that the behavior of the light yield statistic follows the radioactive decay law 
\begin{equation}
S_{ly}  \propto N_{scint}/\exp(-t/t_0),
\end{equation}
where $t_0$ is the time constant for the radiation damage.  
The independent fits to the two periods using this relation are quite good and return $t_0$ $\sim$750 -- 1000~days.  The light yield for blend \#3 is high because of the the PPO mass fraction was high, as seen in Fig.~\ref{fig:PPO_bis_PS}.  This suggests that the light yield from the scintillator could have been increased with more PPO, but there was no more available without additional purchases.

This radiological damage induced by the $\alpha$ source was not seen in the R\&D efforts leading up to scintillator production.  
As discussed, efforts to identify a liquid or solid scintillator standard were considered unsuccessful and there was no time available to restart those investigations.  After the tests of chemical compositon showed the scintillator was blended correctly, the approach taken was to approve blend batches so long the light yield continued to follow the trend seen in Fig.~\ref{fig:trans_lightYield}.

\section{Scintillator Transport}
\label{sect:transport}

Scintillator transport to the far detector at Ash River was managed by Wayne Transport, Rosemont, MN\footnote[18]{http://www.waynetransports.com}.  The scintillator was transported using dedicated, insulated, stainless steel tanker trailers equipped to maintain oil in the temperature range 18.3 -- 23.9~$^\circ$C from filling at Wolf Lake terminals to delivery at Ash River.  The tanker trailers were initially cleaned commercially to transport H1 food grade lubricants, or lubricants used in food processing environments where there is some possibility of incidental food contact\footnote[19]{DIN V 0010517, 2000-08 -- Food Grade Lubricants - Definitions and Requirements}.  All tanker trailers were qualified with a rinse of clean mineral oil that was tested with a tintometer to the mineral oil requirement.  If a tanker trailer did not qualify, it was recleaned and retested until it passed.  Since no loads of scintillator were found to be contaminated, the tanker trailers were not subsequently cleaned.  The tanker trailers were car-sealed at Wolf Lake Terminals after scintillator was loaded and car-sealed at the NOvA far detector building after unloading for the return trip.  At loading, a tintometer test was performed on a sample of the scintillator prior to departure.  At arrival, a tintometer test was performed on a sample of the scintillator prior to unloading.  These additional tests were to assure that the scintillator had not been contaminated during transport.  There were 414 tanker trailer loads shipped to Ash River and there were no rejections.  

There were 9 tanker trailer loads of scintillator shipped from Wolf Lake to Fermilab.  Eight were loaded into the near detector and the ninth is being held as a reserve.  The tanker trailer for these deliveries was owned by Fermilab.  It met the same cleanliness standards as those from Wayne Transports.  Tintometer tests at Wolf Lake and Fermilab were made, as for the shipments to Ash River.  There were no rejections.  

The history of the scintillator deliveries to Ash River and Fermilab are shown in Fig.~\ref{fig:scintTrans-delivery}.
\begin{figure}[h]
\centerline{\includegraphics[width=2.95in,height=2.46in]{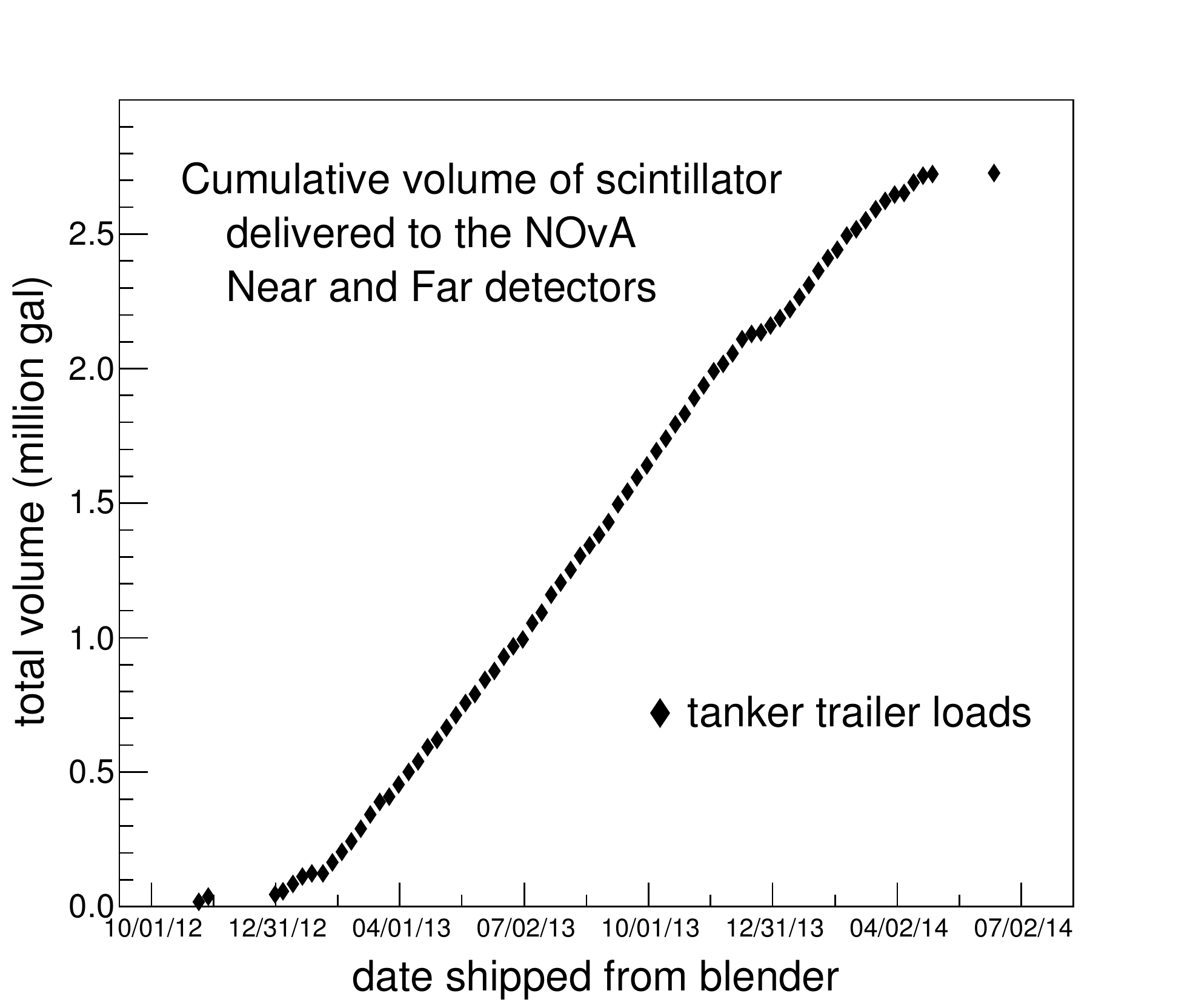}}
\caption{History of the month-by-month liquid scintillator deliveries to the NOvA near and far detectors.  The total volume of liquid scintillator was delivered in 20 months.}
\label{fig:scintTrans-delivery} 
\end{figure}

\section{Summary}
\label{sect:summary}

The NOvA experiment blended and delivered 8.8 kt of liquid scintillator to its near and far detectors as the active detector medium.  The composition of the scintillator was developed to meet the science requirements of the experiment within the cost constraints imposed by its funding profile.  The blending was done commercially at Wolf Lake Terminals in Hammond, IN.  The scintillator was shipped to the detectors at Ash River and Fermilab using dedicated stainless steel tanker trailers that were cleaned to food grade standards.  The production took 20 months to complete.  

A rigorous set of quality control procedures was put in place to verify that the liquid scintillator was blended with components that would achieve scintillator performance requirements.  After blending, the scintillator was tested to verify that it met its transparency, light yield, and conductivity requirements.  The blended scintillator was tested again for transparency before shipping and then again upon arrival at the near and far detectors to assure that the scintillator had not been contaminated during transport.  

\bigskip

\noindent {\bf Acknowledgements} \\

\noindent This work was supported in part by the DOE Office of High Energy Physics through grant DE-SC0010120 to Indiana Universtiy.  Fermi National Accelerator Laboratory (Fermilab) is operated by Fermi Research Alliance, LLC under Contract No. DE-AC02-07CH11359 with the U.S. Department of Energy, Office of Science.  The authors wish to thank the many people who helped make this work possible. At IU: B.~Adams, F.~Busch, C.~Canal, M.~Gebhard, A.~Hansen, T.~Harmon, J.~Musser, E.~Pierson, E.~Steele, R.~Tayloe, and D.~Zipkin;  at Fermilab: E.~Baldina, B. Cibic, K.~Kephart, D.~Pushka, T.J.~Sarlina, R.~Tesarek, and J.~VanGemert; at Renkert Oil:  S.~Kelly and N.~Miller; at Wolf Lake: G.~Calarie, N.~Cave, J.~Hlebek, C.~McClellan, J.~Patton, and E. Sprenne.

\bibliographystyle{elsarticle-num}

\bibliography{scintillator}


\end{document}